\documentclass[published]{JHEP}
\JHEP{09(1999)010}

\usepackage{amsfonts}

\newcommand{\be}{\begin{equation}}
\newcommand{\ee}{\end{equation}}
\newcommand{\bea}{\begin{eqnarray}}
\newcommand{\eea}{\end{eqnarray}}

\newcommand{\U}{{\rm U}}
\newcommand{\SU}{\mathop{\rm SU}}
\newcommand{\nn}{\nonumber}
\newcommand{\lf}{\left}
\newcommand{\ri}{\right}
\newcommand{\f}{\frac}
\newcommand{\dagg}{\dagger}
\def\cL{{\cal L}}
\def\tr           {\mathop{\rm tr}}
\def\ha           {\frac{1}{2}}
\def\al           {\alpha}
\def\bet           {\beta}
\def\ch           {\chi}
\def\chb          {\bar \chi}
\def\del           {\delta}
\def\ep           {\epsilon}
\def\et           {\eta}
\def\Hb            {{\bar H}}
\def\ga           {\gamma}
\def\Ga           {\Gamma}
\def\la           {\lambda}
\def\om           {\omega}
\def\Om           {\Omega}
\def\ph           {\phi}
\def\vph           {\varphi}
\def\ps           {\psi}
\def\rh           {\rho}
\def\si           {\sigma}
\def\ze           {\zeta}
\def\pl           {\partial}
\def\na           {\nabla}
\def\phb           {{\bar{\phi}}}

\title{A $3d$ topological sigma model and D-branes}

\author{Ali Imaanpur\\
        Institute for Studies in Theoretical Physics and Mathematics\\
        P.O. Box 19395-5531, Tehran, Iran, and\\
        Department of Physics, School of Sciences, Tarbiat 
        Modares University,\\
        P.O. Box 14155-4838, Tehran, Iran\\
        Email: \email{aimaanpu@theory.ipm.ac.ir}}

\abstract{A 3d topological sigma model describing maps from a
3-manifold $Y$ to a Calabi-Yau 3-fold $M$ is introduced. As the model
is topological, we can choose an arbitrary metric on $M$.  Upon
scaling up the metric, the path integral by construction localizes on
the moduli space of special lagrangean submanifolds of $M$. We couple
the theory to dynamical gauge fields and discuss the case where $M$
has a mirror and the gauge group is $\U(1)$.}

\received{July 8, 1999}
\accepted{September 9, 1999}

\begin{document}
\section{Introduction}

Two dimensional topological sigma models~\cite{WSIG} have found many
interesting applications both in physics and mathematics.  In string
theory, for example, the effective field theory of curved D3-branes
wrapped around holomorphically embedded 2-cycles in a hyper K\"ahler
manifold, in a limit where the cycle shrinks to zero size, turns out
to be a 2d topological sigma model~\cite{BSV}.  This effective field
theory has then been used to study the existence of bound states of
D-branes.  From the mathematical point of view, topological sigma
models have provided an alternative formulation of Floer groups and
Jones polynomials in knot theory. As in the case of Donaldson theory,
where physical reformulation~\cite{WTFT} has proven to be very useful
and enlightening, specially in the analysis and determination of
Donaldson invariants~\cite{WS, WMO}, one may also hope that the
reformulation of Floer groups and Jones polynomials in terms of a
topological quantum field theory shed light on the structure of these
topological invariants.

The sigma model constructed in~\cite{WSIG} (the A-model~\cite{MIR}) is
a theory of maps \mbox{$X : \Sigma \to M$}, where $\Sigma$ is a
Riemann surface and $M$ a complex manifold. The topological structure
of the model allows one to choose an arbitrary metric on $M$.  When
the metric is scaled up, it can be seen that the dominent
contributions to the path integral come from the holomorphic maps
satisfying
\be
\pl_{\bar z} X_\al =0 \, ,\label{hol}
\ee
where $z, {\bar z}$ are complex coordinates on $\Sigma$, and $\al,
{\bar \al}$ indicate the complex tangent indices on $M$.

Interestingly, the holomorphicity condition~(\ref{hol}) also appears
in the study of D-branes wrapping around Riemann surfaces embedded in
$M$. Roughly speaking, type II string theory compactified on
$\mathbb{R}^6\times M$, $M$ being a hyper K\"ahler manifold, has BPS
states in the form of D3-branes wrapping around $S^1 \times
\Sigma$. $\Sigma$ is embedded in $M$ through $X$, so $X$ is part of
the brane coordinates in the ambient space. One can see that these
solitonic states in general break the supersymmetry of the underlying
superstring theory unless the embedding maps are holomorphic and the
$\U(1)$ connection (being part of the low energy degrees of freedom)
is flat.

Considering D3-branes wrapping around 3-cycles in a Calabi-Yau 3-fold
$M$ and demanding supersymmetry leads to some new constraints known as
the special lagrangean submanifold conditions~\cite{BBS, HL}
 \bea
 X^* k=0\, ,\qquad  X^* ({\rm Im}\, \Om )=0  \label{SL}\, .
 \eea 
 Here $X$ is an embedding of a 3-manifold $Y$ to  
 $M$. $k$ and $\Om$ represent the K\"ahler and holomorphic 3-form on $M$,  
 respectively, and $*$ indicates the pull-back operation. The 3-cycles 
satisfying eqs.~(\ref{SL}) are  sometimes called supersymmetric 
3-cycles. 

Now taking eqs.~(\ref{SL}) as the corresponding equations 
to~(\ref{hol}) in 3 dimensions, it is interesting to see whether the
above 2-dimensional sigma model can be generalized to 3 dimensions. If
such, then the path integral of the corresponding sigma model, in some
scaling limit of the parameters entering the lagrangean, would
localize on the moduli space of solutions to~(\ref{SL}).  As in two
dimensions, one may expect that the effective low energy description
of curved D5-branes wrapping around $S^1 \times S^1 \times Y$, where
$Y$ is a supersymmetric 3-cycle in $M$ shrinking to zero size, is
given by a 3d topological sigma model of above type.  In~\cite{DAVIS},
using the BV method, a 3 dimensional sigma model has been constructed
which has some common properties with the one that we will introduce
here. Our motivation, approach, and results, though, are different.
 
This paper is organized as follows. To begin with, in section~\ref{s2}
we take equations~(\ref{SL}) as the starting point for the
construction of a 3-dimensional topological sigma model. Apart from
the bosonic fields $X$, we introduce some fermionic fields needed to
define a BRST-like symmetry. Having had the equations and BRST
symmetry, we proceed to construct a lagrangean which is BRST
trivial. Next we discuss the moduli space of solutions to~(\ref{SL})
where the dominent contributions to the path integral come from. In
section~\ref{s3}, we couple the theory to dynamical gauge fields. As
in~\cite{WSIG}, a consistent coupling to the gauge fields requires a
modification of the BRST transformation rules.  The lagrangean
describing the dynamics of gauge multiplet is then obtained by
dimensional reduction of the lagrangean of twisted $N=2$ SYM in four
dimensions. The case where the gauge group is $\U(1)$ and the target
manifold $M$ has a mirror ${\widetilde M}$ is of particular
interest. The moduli space of solutions to the fixed point equations,
in this case, turns out to be parametrized by the mirror manifold
${\widetilde M}$. At the end, we have collected the conventions and
some derivations in an appendix.

\section{A topological sigma model}\label{s2}

In this section we aim to construct a topological sigma model of maps
$X: Y\to M$, which has the special lagrangean conditions~(\ref{SL}) as
its fixed point equations. Let us indicate the indices on $Y$ by $i,
j, \ldots,$ and those on $M$ by $\mu ,\nu ,\ldots$ It will prove
convenient, if instead of~(\ref{SL}), we consider the equations
$s={\sf k}_{ij}=0$, with
 \bea
 s&=&1 +\f{1}{3!\sqrt h}\ep^{ijk}X^\mu_i  X^\nu_j  X^\la_k \Om_{\mu\nu\la}\,,\nn \\
{\sf k}_{ij}&=& X^\mu_i X^\nu_j k_{\mu\nu}\nn \, ,
 \eea 
where we have defined $X^\mu_i \equiv \pl_i X^\mu$, and $h$ 
 indicates the determinant of the induced metric; 
 \[
 h =\det h_{ij}= 
 \f{1}{3!}\{\ep^{ijk}\ep^{mnl}h_{im}h_{jn}h_{kl}\}= 
 \det (\pl_i X^\mu \pl_j X^\nu g_{\mu\nu})\, .
 \]
As we are describing the theory on the embedded 3-manifold, in the
following, we always use the induced metric to raise or lower the
indices on $Y$.  The prescription for the construction of the
lagrangean is now as follows.

First we introduce a  
ghost field $\xi^\mu$, the fermionic partner of $X^\mu$, and the 
BRST operator $\del$ with an action
 \be
 \del X^\mu =i\ep\, \xi^\mu \, , \qquad \del \xi^\mu =0\, ,\label{tr1}
 \ee
where $\ep$ is a constant anticommuting parameter. $\xi^\mu$ is a
section of $X^*(T)$, with $T$ the tangent bundle of $M$. Further we
need to introduce an anti-ghost field $\ch$, an anti-ghost two-form
$\rh_{ij}$ (the conjugate fields to $s$ and ${\sf k}_{ij}$), as well
as a scalar field $H$ and a two-form $H_{ij}$. The transformation laws
are now defined to be
 \bea
  \del (h^{1/4}\ch) &=&\ep\, h^{1/4} H \,,\qquad\del (h^{1/4}H) =0\,, \nn \\ 
  \del (h^{1/4}\rh_i^{\ j}) &=&\ep\, h^{1/4}H_i^{\ j}\, ,\qquad
 \del (h^{1/4}H_i^{\ j})=0 \, .
 \label{tr2}
 \eea 
Let us define the operator $Q$ by $\del \Phi =-i\ep \{Q , \Phi\}$, for
any field $\Phi$.  We would like the lagrangean to be a BRST
commutator, i.e.\  $\cL= i\{Q , \Psi\}$, for some gauge fermion $\Psi$.
A minimal choice for $\Psi$ is
 \be
 \Psi = \f{1}{4\la}{\sqrt h}\left (\chb H + \ch \Hb + \rh_{ij}H^{ij} -2 \chb s 
 -2\ch {\bar s} 
 -2 \rh_{ij}{\sf k}^{ij}\right) ,  \label{PSI}
 \ee
 where $\la$ is an arbitrary real parameter.
 The action now reads 
 \bea
 S = i\int \, \{Q ,  \Psi \} &=& 
  \f{1}{2\la}\int {\sqrt h}\,  d^3\si \Biggl[ -H\Hb +\Hb s +H{\bar s}- 
 \f{1}{2}H_{ij}H^{ij} +H_{ij}{\sf k}^{ij} -
\nn \\&&\hphantom{\f{1}{2\la}\int {\sqrt h}\,  d^3\si \Biggl[}
-  i h^{-1/4}\chb \f{\del (h^{1/4}s)}{\del X^\rh}\xi^\rh 
 -i h^{-1/4}\ch \f{\del (h^{1/4}{\bar s})}{\del X^\rh}\xi^\rh -
\nonumber\\&&\hphantom{\f{1}{2\la}\int {\sqrt h}\,  d^3\si \Biggl[}
 -i h^{-1/4}\rh_i^{\ j}
 \f{\del (h^{1/4}h^{ik}{\sf k}_{kj})}{\del X^\rh}\xi^\rh \Biggr] \nn \, .  
 \eea 

The auxiliary fields $H$ and $H_{ij}$ have no dynamics and can be
integrated out using their equations of motion. Doing this, the
bosonic part of the action becomes
\[
\f{1}{\la}\int {\sqrt h}\,  d^3\si\,  \lf( \f{1}{2} s{\bar s} + \f{1}{4}
{\sf k}^{ij}{\sf k}_{ij} \ri)  .
\]
However, in the appendix we show that
 \be
 \f{1}{2}s{\bar s} + \f{1}{4}{\sf k}^{ij}{\sf k}_{ij}= 
 1 +\f{1}{2\cdot 3!\sqrt h}\ep^{ijk} X^\mu_i  
X^\nu_j  X^\la_k (\Om_{\mu\nu\la} 
 + \Om^\dagg _{\mu\nu\la})\, . \label{SAT}
 \ee
Therefore, using the above identity, we can write the action as
 \bea
 S &=& \f{1}{\la}\int d^3\si {\sqrt h}\lf[ 
 1+ \f{1}{2\cdot 3!{\sqrt h}}\ep^{ijk} 
 X^\mu_i X^\nu_j X^\la_k(\Om_{\mu\nu\la}+\Om^\dagger_{\mu\nu\la})\ri.- \nn\\[3pt]
& &\quad
- \f{i}{4}g_{\mu\nu}h^{ij}X^\mu_i\,  \chb \na_j\xi^\nu\left(1   
 + \f{1}{3!{\sqrt h}}\ep^{mnl}X^\rh_m X^\si_n X^\del_l\Om_{\rh\si\del}\right) 
 + {\rm h.c.}- \nn \\[3pt]
& &\quad - \f{i}{4{\sqrt h}}\ep^{ijk} X^\nu_j X^\la_k\Om_{\mu\nu\la}\, \chb 
 \pl_i\xi^\mu  
 - \f{i}{2\cdot 3!{\sqrt h}}\ep^{ijk}X^\mu_i X^\nu_j X^\la_k \partial _\rh 
 \Om_{\mu\nu\la}\chb \xi^\rh    + {\rm h.c.}-\nn \\  [3pt]
 &&\quad - \f{i}{4}h^{ij}g_{\mu\nu}X^\mu_i{\sf k}_{mn}\rh^{mn} 
 \na_j\xi^\nu   
+i g_{\mu\nu}\rh^{ik}{\sf k}^{j}_{\ k}X^\mu_i \na_j \xi^\nu-
\nn \\ [3pt]
& &\quad - \lf. i k_{\mu\nu}X^\mu_i \rh^{ij}\pl_j\xi^\nu 
 -\f{i}{2} \partial_\rh k_{\mu\nu}X^\mu_i X^\nu_j \rh^{ij}\xi^\rh \ri]
 \label{obtain},
 \eea
where $\na_i\xi^\mu =\pl_i\xi^\mu +\pl_iX^\nu  \Ga_{\nu\la}^\mu \xi^\la$.

Since the K\"ahler form is closed, the last two terms combine to  
 \bea
\lefteqn{ \rh^{ij}\Bigl(\pl_i\xi^\mu X^\nu_j k_{\mu\nu} +\pl_j\xi^\nu X^\mu_i k_{\mu\nu} 
 +X^\mu_i X^\nu_j \pl_\rh k_{\mu\nu}\xi^\rh\Bigr)=}
 \nn\\[3pt]
 &&\qquad= \rh^{ij}\Bigl(\pl_i\xi^\mu X_j^\nu k_{\mu\nu} + \pl_j\xi^\nu X^\mu_i k_{\mu\nu} 
 +\pl_i k_{\rh\nu}X^\nu_j\xi^\rh +\pl_jk_{\mu\rh}X^\mu_i\xi^\rh\Bigr)\nn\\[3pt]
 &&\qquad= \rh^{ij}\Bigl(\pl_i(X^\nu_j\xi^\mu k_{\mu\nu}) - \pl_j(X^\nu_i\xi^\mu 
 k_{\mu\nu})\Bigr)
 \, . \label{omega}
 \eea 
 Likewise, as the holomorphic 3-form is closed, the terms  
 in the third line can be 
 written as
 \bea
 \lefteqn{\f{1}{2{\sqrt h}}\ep^{ijk}\chb\left\{ \pl_i\xi^\mu X^\nu_j X^\la_k\Om_{\mu\nu\la}
 +\f{1}{3}X^\mu_iX^\nu_jX^\la_k \pl_\rh\Om_{\mu\nu\la}\xi^\rh \right\}=} \nn\\[3pt]
 &=& \f{1}{3!{\sqrt h}}\ep^{ijk}\chb\left\{ \pl_{[i}\xi^\mu X^\nu_jX^\la_{k]}
  \Om_{\mu\nu\la} \!+\xi^\rh \Bigl( X^\nu_jX^\la_k\pl_i\Om_{\rh\nu\la} 
 \!+ X^\mu_iX^\la_k\pl_j\Om_{\mu\rh\la}
 \!+ X^\mu_iX^\nu_j\pl_k\Om_{\mu\nu\rh}\Bigr) \right\} \nn \\[3pt]
 &=&\f{1}{2{\sqrt h}}\ep^{ijk}\chb\, \pl_i(\xi^\mu X^\nu_jX^\la_k\Om_{\mu\nu\la})
 \label{Omega1} \, .
 \eea
\pagebreak[3] 

\noindent Let $\ch =\varrho + i\ze$, where $\varrho$ and $\ze$ are real fields,  
the action now reads
 \bea
 S &=& \f{1}{\la}\int d^3\si {\sqrt h}
 \lf[ 1+ \f{1}{2\cdot 3!{\sqrt h}}\ep^{ijk} 
 X^\mu_i X^\nu_j X^\la_k(\Om_{\mu\nu\la}+\Om^\dagger_{\mu\nu\la})-\ri. \nn\\
 &&- \f{i}{2} \lf(1  + \f{1}{2\cdot 3!{\sqrt h}}
\ep^{mnl}X^\rh_m X^\si_n X^\del_l
 (\Om_{\rh\si\del}+ \Om^\dagger_{\rh\si\del})\ri)
 g_{\mu\nu}h^{ij}X^\mu_i\, \varrho \na_j\xi^\nu -  \nn \\
 &&-\f{1}{4\cdot 3!{\sqrt h}}\ep^{mnl}X^\rh_m X^\si_n X^\del_l
 (\Om_{\rh\si\del}- \Om^\dagger_{\rh\si\del}) g_{\mu\nu}h^{ij}X^\mu_i\, 
 \ze \na_j\xi^\nu- \nn \\
 &&- \f{i}{4{\sqrt h}}\ep^{ijk}\lf( 
 \varrho\,  \pl_i\lf(\xi^\mu X^\nu_j X^\la_k 
 (\Om_{\mu\nu\la}+\Om^\dagger_{\mu\nu\la})\ri) 
 -i\ze\, \pl_i\lf(\xi^\mu X^\nu_j X^\la_k 
 (\Om_{\mu\nu\la}-\Om^\dagger_{\mu\nu\la})\ri)\ri) -
 \nn \\
 &&- \lf.  \f{i}{4}h^{ij}g_{\mu\nu}X^\mu_i{\sf k}_{mn}\rh^{mn} 
 \na_j\xi^\nu 
+i g_{\mu\nu}\rh^{ik}{\sf k}^{j}_{\ k}X^\mu_i \na_j \xi^\nu 
 -i \rh^{ij}\pl_i(X^\nu_j \xi^\mu k_{\mu\nu}) \ri] .\label{action2}
 \eea

Recall that the action was defined to be a BRST commutator.  Therefore
the path integral -- upon assuming that the measure is also invariant
under the BRST transformations -- is independent of the parameter
$\la$ as well as the metric on $M$ (or more generally independent of
any data entering $\Psi$).  Hence, in computing the path integral, an
arbitrary and convenient value for these parameters can be chosen.
For instance, we may take the limit $\la \to 0 $. Or equivalently, we
may scale the metric as $g_{\mu\nu}\to tg_{\mu\nu}$ for $t$ a real
parameter, and then take the limit $t\to \infty $.  In both limits,
the path integral will localize around the solutions of the following
equations \[ {\sqrt h} + \f{1}{2\cdot 3!}\ep^{ijk} X^\mu_i X^\nu_j
X^\la_k(\Om_{\mu\nu\la}+\Om^\dagger_{\mu\nu\la})=0 \, .  \] By
eq.~(\ref{SAT}), this is equivalent to eqs.~(\ref{SL}) defining the
special lagrangean submanifolds of $M$. Let us indicate the moduli
space of solutions to these equations by ${\cal M}_{sl}$ and study the
tangent space at $Y$.

\subsection{The tangent space to ${\cal M}_{sl}$}

Let $X^\mu$ describe a special lagrangean submanifold of $M$.  We
deform $X^\mu$ to a nearby map $X^\mu + \del X^\mu$, and ask under
what conditions this new map is still a special lagrangean
submanifold. For $\del X^\mu$ not to have any component obtainable by
a diffeomorphism on $Y$, we demand that $\del X^\mu$ belong to the
normal bundle at $Y$, i.e. 
\[ 
g_{\mu\nu}\pl_iX^\mu \del X^\nu =0 \, .
\]
  
To find the tangent space to ${\cal M}_{sl}$ at $Y$, we impose the
special lagrangean conditions (${\rm Im}\, s={\sf k}_{ij}=0$) on
$X^\mu +\del X^\mu$. Following the same line of arguments as
in~(\ref{omega}), for the lagrangean condition (i.e.\ ${\sf k}_{ij}=0$) we get
 \be
 0= \pl_{[i}(\del X^\mu)X^\nu_{j]}k_{\mu\nu}=
 \pl_i(X^\nu_j\del X^\mu k_{\mu\nu})- \pl_j(X^\nu_i\del X^\mu k_{\mu\nu})
 \equiv \pl_i\om_j - \pl_j \om_i \, ,
 \label{dom}
 \ee
 where $\om_i \equiv \del X^\mu X^\nu_i k_{\mu\nu}$. As the pull-back of the 
K\"ahelr form is zero, the complex structure provides an isomorphism between 
the tangent space, $T_YM$, and the normal space, $N_YM$ at $Y$. As such, 
$J^\mu_{\ \nu}\del X^\nu$ is a tangent vector at $Y$. Contracting this vector 
with the metric and pulling it back gives a one-form on $Y$. Hence, $\om_i$ 
is a one-form on $Y$.

The implication of the special lagrangean condition can be seen more 
easily  by writing  
 the K\"ahler form and $\Om$ in the vielbein 
 bases (here $n=3$),
 \bea
  k&=& E^1\wedge E^{n+1} + E^2\wedge E^{n+2} + E^3\wedge E^{n+3}\,, \nn \\
 \Om &=& (E^1+iE^{n+1})\wedge (E^2+iE^{n+2})\wedge(E^3+iE^{n+3}) \nn
 \, ,  
 \eea
 or in components
 \bea
  k_{\mu\nu}&=&E^a_\mu E^{n+a}_\nu -E^a_\nu E^{n+a}_\mu \,,\nn \\
 \Om_{\mu\nu\la}&=&\ep_{abc}(E^a+iE^{n+a})_\mu (E^b+iE^{n+b})_\nu
 (E^c+iE^{n+c})_\la  \label{com}\, ,
 \eea
 where $a,b,\ldots=1,2,3$. We adopt the vielbeins such that $E^a_\mu$ 
and $E^{n+a}_\mu$ span the tangent and normal 
spaces at $Y$, respectively. The pull-back of $E^a_\mu$ is then  
naturally to be the vielbeins $e^a_i$ on $Y$ 
 \be
 e^a_i = X^\mu_i E^a_\mu \, ,\qquad X^\mu_iE^{a+n}_\mu=0 \, ,\qquad  
 \del X^\mu E^a_\mu=0 \, . \label{adopt}
 \ee
 If we impose the second condition ${\rm Im}\, s = \f{1}{3!{\sqrt h}}
 \ep^{ijk}X^\mu_i X^\nu_j 
 X^\la_k {\rm Im}\, \Om_{\mu\nu\la}=0$ on $X^\mu +\del X^\mu$
 we get
 \bea
 0 &=& \ep^{ijk}{\rm Im}\left \{ \pl_i(\del X^\mu) X^\nu_j X^\la_k\Om_{\mu\nu\la}
 +\f{1}{3}X^\mu_iX^\nu_jX^\la_k \pl_\rh\Om_{\mu\nu\la}\del X^\rh\right \} \nn\\
 &=&\ep^{ijk}{\rm Im}\, \pl_i(\del X^\mu X^\nu_jX^\la_k\Om_{\mu\nu\la})
 \label{Omega} \, .
 \eea
 using~(\ref{com}) and~(\ref{adopt}), this can be written as
 \bea
 0 &=& \pl_i(\ep^{ijk}\ep_{abc}\del X^\mu E^{n+a}_\mu X^\nu_jE^b_\nu 
 X^\la_kE^c_\la ) \nn \\
 &=& \pl_i(\ep^{ijk}\ep_{abc}\del X^\mu E^{n+a}_\mu e^b_j e^c_k)\nn \\
 &=& \pl_i(\ep^{ijk}\ep_{ljk}\del X^\mu E^{n+a}_\mu e^l_a) \nn\\
 &=&2 \pl_i(\del X^\mu E^{n+a}_\mu e^i_a)\nn \\
 &=&2 \pl_i(\del X^\mu E^{n+a}_\mu E^\nu_a X^i_\nu) \nn \\
 &=&2 \pl_i(\del X^\mu X^{i\nu}k_{\mu\nu})
 =2\pl_i \om^i \label{daom}\, .
 \eea

Equations~(\ref{dom}) and~(\ref{daom}) imply that for the map $X^\mu +
\del X^\mu$ to be a special lagrangean submanifold of $M$, the
one-form $\om_i= \del X^\mu X^\nu_i k_{\mu\nu}$ must be closed and
coclosed~\cite{MC, SYZ}.  Therefore any harmonic one-form $\om$ on $Y$
specifies a direction in which a special lagrangean submanifold can be
deformed in ${\cal M}_{sl}$, and the dimension of ${\cal M}_{sl}$
equals $b_1$, the first Betti number of the manifold $Y$. Looking back
to~(\ref{action2}) shows that, in the background of a map $X^\mu$
describing a special lagrangean submanifold of $M$, zero modes of
$\xi^\mu$ satisfy the same equations that $\del X^\mu$ does
in~(\ref{dom}) and~(\ref{daom}). Consequently, if $b_1 > 0$, there
exist $\xi^\mu$ zero modes and the partition function identically
vanishes.  To soak up the zero modes, while maintaining the
topological characteristic of the theory, we need to insert some BRST
invariant operators with the right ghost number in to the path
integral.  This leads us to look for the cohomology classes of the
operator $Q$. Following~\cite{WSIG}, these cohomology classes can be
constructed and be shown that are in correspondence with the de Rham
cohomology classes of~$M$.

\section{Coupling to gauge fields}\label{s3}

We would like now to couple the sigma model of the previous section to
dynamical gauge fields of a compact gauge group $G$.  As
in~\cite{WSIG}, $G$ is taken to be the group of automorphisms of $M$
preserving all the structures on the manifold. The $G$ action on $M$
is through the vector fields $V_a= V_a^\mu e_\mu$, $a=1,\ldots ,n$
($n$ is the dimension of $G$), representing the group generators,
i.e.
\[
 [V_a , V_b] =f_{ab}^{\ \ c} V_c \, ,
 \]
 for $f_{abc}$ the structure constants of $G$. The fact 
that the $G$ action preserves all the structures on $M$ means 
 that the Lie derivatives of the metric, K\"ahler form, and holomorphic 
 3-form all have to vanish:
 \[
 \cL_{V_a}(g)=\cL_{V_a}(k)=\cL_{V_a}(\Om )=0\, ,
 \]
 implying that
 \bea
 \na_\mu V_{a\nu} + \na_\nu V_{a\mu} &=&0 \nn \\
 k_{\rh\nu}\na_\mu V^\rh_a + k_{\mu\rh}\na_\nu V^\rh_a &=&0 \nn \\
\Om_{\mu\nu\rh}\na_\la V^\rh_a + \Om_{\mu\rh\la}\na_\nu V^\rh_a 
 + \Om_{\rh\nu\la}\na_\mu V^\rh_a &=&0 \nn \, .
 \eea
In the following, first we discuss the lagrangean of the gauge sector.
Next, as the gauge group acts nontrivially on $M$ and since $\del^2$
acting on the gauge multiplet produces a gauge transformation, we will
see how the transformation rules in~(\ref{tr1}) need to be modified.

The topological lagrangean describing the dynamics of gauge sector can
be obtained more conveniently by dimensional reduction of
Donaldson-Witten theory in four dimensions.\footnote{Alternatively,
one may follow~\cite{BAU} to construct the lagrangean of the gauge
multiplet.} The lagrangean of the Donaldson-Witten theory, on the
other hand, can be obtained by twisting the lagrangean of $N=2$ SYM
theory on $\mathbb{R}^4$.  As a result, one finds the following field
content and lagrangean.  In the bosonic sector there are a gauge field
$A_\mu$, two scalars $\ph$ and $\phb$, and in the fermionic sector we
have a one-form $\ps_\mu$, a scalar $\et$, and a self-dual two-form
$\ch_{\mu\nu}$.  The lagrangean on an arbitrary smooth four manifold
then reads~\cite{WTFT}
 \bea
 \cL &=& \tr\lf[\f{1}{4}F^+_{\mu\nu}F^{+\mu\nu}+\f{1}{2}\ph D_\mu D^\mu\phb 
 -i\et D_\mu\psi^\mu +iD_\mu\ps_\nu \ch^{\mu\nu}-\ri. 
\nn \\&&\hphantom{\tr\Biggl[}
-\lf. \f{i}{8}\ph[\ch_{\mu\nu} , \ch^{\mu\nu}] 
 -\f{i}{2}\phb[\ps_\mu , \ps^\mu ]
 -\f{i}{2}\ph [\et , \et]-\f{1}{8}[\ph , \phb ]^2 \ri]  .\label{witten}
 \eea
 This action is invariant under the transformations
 \begin{equation}
\begin{array}[b]{rclcrclcrcl}
 \del A_\mu &=& i\ep\, \ps_\mu \,,&\qquad& \del\ph &=&0\, 
 ,&\qquad& \del\phb &=&2i\ep\, \et \, ,
\\ 
 \del\et &=& \displaystyle\f{1}{2}\ep[\ph , \phb ]\, ,&\qquad&
 \del\ps_\mu&=& -\ep\, D_\mu\ph\,,&\qquad& 
 \del\ch_{\mu\nu}&=& 2\ep\, F_{\mu\nu}^+\, .
 \label{evi}
\end{array}
 \end{equation}

Take the underlying four manifold to be $Y\times \mathbb{R}$, for $Y$
the embedded 3-manifold.  Let $t$ parametrize the line $\mathbb{R}$
and simply require that the fields not depend on $t$. Setting
\mbox{$\vph =A_0\, , \ a= \ps_0 \, ,\ \ch_i= \ch_{i0}=\ha
\ep_{ijk}\ch^{jk}$}, the lagrangean describing the dynamics of gauge
multiplet on $Y$ reads
\bea
 \cL_{G} &=& \tr\lf[\f{1}{4}F_{ij}F^{ij} +\f{1}{2}\ep_{ijk}F^{ij}D^k\vph 
 -\f{1}{2} D_i\vph D^i\vph 
 +\f{1}{2}\ph D_i D^i\phb 
 -i\et D_i\psi^i - \ri. 
\nn \\&&\hphantom{\tr\Biggl[}
- i\ch_iD^ia -i \ep_{ijk}\ch^iD^j\ps^k -[\vph , \ps_i]\ch^i 
 -\f{i}{2}\ph[\ch_{i} , \ch^{i}] 
 -\f{i}{2}\phb[\ps_i , \ps^i ]-
 \nn \\&&\hphantom{\tr\Biggl[}
 -\f{i}{2}\phb[a , a ] +\lf. \et[\vph , a]+ \f{1}{2}[\vph , \ph][\vph , \phb]
 -\f{i}{2}\ph [\et , \et]-\f{1}{8}[\ph , \phb ]^2 \ri]\, .
\label{monopole}
 \eea
 The symmetry transformations of this action are obviously read 
from~(\ref{evi}) to be
 \begin{equation}
\begin{array}[b]{rclcrclcrcl}
 \del A_i &=& i\ep\, \ps_i \,,&\qquad& \del\phb &=&2i\ep\, \et \, , 
 &\quad& \del\et &=&\displaystyle \f{1}{2}\ep [\ph , \phb ]\, , 
\\
 \del \vph &=& i\ep\, a \,,&\qquad& \del a &=&-\ep [\vph , \ph]\, , 
 &\quad& \del\ph &=&0\, , 
 \nn \\
 \del\ps_i&=& -\ep\, D_i\ph\,,&\qquad& \del\ch_{i}&=&
\displaystyle\ep \left( \f{1}{2}\ep_{ijk}F^{jk}
 +D_i\vph\right) .\label{tr3}
\end{array}
 \end{equation}
 As $Y$ has no boundary, the second term in the action vanishes by 
the Bianchi identity. So   
 the fixed point equations reduce to 
 \[
 F_{ij}=0\, ,\qquad D_i\ph =D_i\vph =0\, .
 \]

Notice that $\del_G\equiv \del^2$ acting on the fields in~(\ref{tr3})
produces an infinitesimal gauge transformation. So to couple the gauge
multiplet to the sigma model we need to change the transformation
rules in~(\ref{tr1}) such that $\del^2$ is not zero but a gauge
transformation. Firstly, note that the infinitesimal action of $G$ on
the coordinates is $\del_G X^\mu \sim \ph^a V^\mu_a$, for $\ph^a$ the
gauge parameter. So to have $\del^2 = \del_G$, we need to change the
transformations~(\ref{tr1}) to
 \[
 \del X^\mu =i\ep\, \xi^\mu\, , \qquad \del\xi^\mu = \ep\, \ph^a V_a^\mu\, .
 \]
This, in particular, gives the proper gauge transformation of $\xi^\mu$ as 
a section of $X^*(T)$:
 \[
 \del_G \xi^\mu \sim \ph^a \pl_\nu V^\mu_a \xi^\nu \, .
\]
With this change, the covariant derivative of $X^\mu$, $D_i X^\mu =
\pl_i X^\mu + A_i^aV_a^\mu$, transforms (just like $\xi^\mu$) as a
vector under $\del_G$.  As we are interested in having a set of gauge
invariant equations, and since the metric, $k$ and $\Om$ are all
invariant under the $G$ action, we replace $s$ and ${\sf k}_{ij}$ with
\bea
 && s= 1 + \f{1}{3!}(\det (g_{\mu\nu}D_iX^\mu D_jX^\nu))^{-1/2}
 \ep^{ijk}D_iX^\mu D_j X^\nu D_k X^\la \Om_{\mu\nu\la}\,,\nn \\
 && {\sf k}_{ij}= D_iX^\mu D_j X^\nu k_{\mu\nu}\nn \, .
 \eea

 Let us now see if any changes need to be made in the transformation 
 rules of $\ch$ and $\rh_{ij}$. Since the action is BRST-exact and 
 we want to maintain its invariance under $\del$, $\Psi$  has to be 
 a gauge invariant quantity
 \[
 0= \del S =\del^2 \Psi = \del_G \Psi\, .
 \]
 On the other hand, as $s$ and 
 ${\sf k}$ are gauge invariant, a look back to~(\ref{PSI}) shows that 
 the conjugate fields $\ch$ and $\rh_{ij}$ 
 also have to be gauge invariant. This implies that the transformation 
 rules~(\ref{tr2}) do not need any changes. The gauge invariant version 
 of the sigma model action can now be derived by varying the $\Psi$ with 
 respect to the newly defined $\del$, $\cL_{\rm sig} = i\{ Q , \Psi \}$. 
This turns out to be  
 the action in 
~(\ref{obtain}), if we do the following substituations: 
 \bea
 \pl_iX^\mu &\to& D_i X^\mu\,, \nn \\
 \pl_i\xi^\mu &\to& \pl_i\xi^\mu + A_i^a \pl_\nu V_a^\mu \xi^\nu \, .
 \eea 
The total lagrangean is then the sum of the lagrangeans 
$\cL_{\rm sig}$ and~(\ref{monopole})
\[
\cL = \cL_G + \cL_{\rm sig} \, .
\]

\subsection{The $\U(1)$ case}

In the case of $\U(1)$ gauge group with a trivial action on $M$, the
fixed point equations constrain the map $X^\mu$ to be a special
lagrangean submanifold, and the $\U(1)$ gauge connection living on the
embedded 3-manifold to be flat. In the previous section, we saw that
${\cal M}_{sl}$ is parametrized by a torus $T^{b_1}$. It is also easy
to see that the moduli space of flat $\U(1)$ connections on $Y$ is
parametrized by a torus $T^{b_1}$. So the dimension of the total
moduli space ${\cal M}$ is $2b_1$. Suppose $M$ has a mirror
${\widetilde M}$.  Quantum mirror symmetry then implies that the total
moduli space ${\cal M}$ is nothing but the mirror manifold
${\widetilde M}$, and $b_1=3$ for the dimensions to match~\cite{SYZ}.
\pagebreak[3]

We conclude that for the $\U(1)$ gauge group, the path integral
calculation of the corresponding topological sigma model reduces to an
integral over the mirror manifold. The lagrangean of such a
topological sigma model coupled to gauge fields should in principle be
obtainable starting from the lagrangean of super Yang-Mills theory on
a Calabi-Yau 3-fold~\cite{JAJ} and then reducing it on to a $3d$
submanifold. We hope to return to this point in future.

\appendix
\section{Conventions and the proof of eq.~(\protect\ref{SAT})}

In this appendix, using the conventions of~\cite{JAJ}, we derive 
equation~(\ref{SAT}).  

Let $M$ be a Calabi-Yau 3-fold. As such, there exists a spinor
$\theta$ on $M$ which is a singlet under the holonomy group
$\SU(3)$. Take $\theta$ to be left-handed and normalize it such that
$\theta^\dagger\theta =1$.  Let us choose a representation for the
gamma matrices such that they are hermitian and antisymmetric;
$\ga_\mu^\dagger =\ga_\mu$, $\ga_{\mu\nu}^\dagger =- \ga_{\mu\nu}$,
$\ga_{\mu\nu\la}^\dagger = -\ga_{\mu\nu\la}$ (where
$\ga_{\mu\nu\la}= \ha\{\ga_\mu , \ga_{\nu\la}\}$). Thus
 \begin{equation}
\begin{array}[b]{rclcrcl}
 \ga_7 \theta&=&-\theta \, ,&\qquad&  \ga_7 \theta^* &=&\theta^* \\
 \theta^\dagger\ga_7 &=&-\theta^\dagger \, ,&\qquad& \theta^t\ga_7 &=&\theta^t \,  ,
\end{array}
 \end{equation}
 $\theta$ and $\theta^\dagger$ are left-handed while $\theta^*$ and $\theta^t$ 
 are right-handed spinors. This in particular implies that 
 \[
 \theta^\dagger\theta^* =\theta^\dagger\ga_{\mu\nu\la}\theta =0 \nn \, . 
 \] 
Also, it is easy to see that $\theta^\dagger \ga_\mu \theta^* =0$. 
 We define 
 \bea
  k_{\mu\nu} &=&i\theta^\dagg \ga_{\mu\nu}\theta 
\nonumber \\\label{k}
  \Om_{\mu\nu\la}&=&\theta^\dagg \ga_{\mu\nu\la}\theta^*\, .
 \eea
As $\theta$ is a singlet under the holonomy group, these  
are nowhere vanishing closed forms. Therefore we recognize $k$ and 
$\Om$ as the K\"ahler and holomorphic 3-form on $M$, respectively. 

To derive eq.~(\ref{SAT}), first note that
 \bea
 \Om_{\mu\nu\la}\Om^\dagg _{\rh\si\del}&=&  
-(\theta^\dagg\ga_{\mu\nu\la}\theta^*)(\theta^t
 \ga_{\rh\si\del}\theta)\nn \\
 &=& -(\theta^\dagg\ga_{\mu\nu\la})\left[\f{1}{2}(1+\ga_7)
- \f{1}{2}\ga^\et\theta\theta^\dagg\ga_\et\right]
 (\ga_{\rh\si\del}\theta)\nn \\
 &=& -\theta^\dagg \ga_{\mu\nu\la}\ga_{\rh\si\del}\theta 
+ \f{1}{2}(\theta^\dagg\ga_{\mu\nu\la}
 \ga^\et\theta)(\theta^\dagg\ga_\et\ga_{\rh\si\del}\theta)   \label{OM} \, ,
 \eea
 where use has been made of the Fierz identity
 \[
 \theta^*\theta^t +\f{1}{2}\ga^\la\theta\theta^\dagger\ga_\la = \f{1}{2}(1+\ga_7 )\, .
 \]
 Using the identity 
 \bea
 \ga_{\rh\si\del}\ga^{\mu\nu\la} &=& \f{i}{2}\ep^{\mu\nu\la}_{\ \ \ \
 \rh\al\bet}
\left(-\f{i}{2}
 \ep^{\al\bet}_{\ \ \ \si\del\al '\bet '}\ga^{\al '\bet '}- \del_{[\si}^{\ [\al}\ga_{\del]}^{\ \bet]}\ga_7 
 -\del_{[\del}^{\ [\bet}\del_{\si]}^{\ \al]}\ga_7 \right)+ \nn \\
&& + \del_\rh^{\ \mu}\left(-\f{i}{2}\ep^{\nu\la}_{\ \ \ \si\del\al '\bet '}\ga^{\al '\bet '}\ga_7 
 - \del_{[\si}^{\ [\nu}\ga_{\del]}^{\ \la]} -\del_{[\del}^{\ [\la}\del_{\si]}^{\ \nu]}\right)+ \nn \\
&& + \del_\rh^{\ \la}\left(-\f{i}{2}\ep^{\mu\nu}_{\ \ \, \si\del\al '\bet '}\ga^{\al '\bet '}\ga_7 
 - \del_{[\si}^{\ [\mu}\ga_{\del]}^{\ \nu]} -\del_{[\del}^{\ [\nu}\del_{\si]}^{\ \mu]}\right) +\nn \\
&& + \del_\rh^{\ \nu}\left(-\f{i}{2}\ep^{\la\mu}_{\ \ \, \si\del\al '\bet '}\ga^{\al '\bet '}\ga_7 
 - \del_{[\si}^{\ [\la}\ga_{\del]}^{\ \mu]} -\del_{[\del}^{\ [\mu}\del_{\si]}^{\ \la]}\right)+ \nn \\
&& + g_{\si\rh}\left(\f{i}{2}\ep^{\mu\nu\la}_{\ \ \ \ \del\al '\bet '}\ga^{\al '\bet '}\ga_7 
 + \del_\del^{\ \mu}\ga^{\nu\la}+ \del_\del^{\ \la}\ga^{\mu\nu}+ \del_\del^{\ \nu}\ga^{\la\mu} \right)- \nn \\
&& - g_{\del\rh}\left(\f{i}{2}\ep^{\mu\nu\la}_{\ \ \ \ \si\al '\bet '}\ga^{\al '\bet '}\ga_7 
 + \del_\si^{\ \mu}\ga^{\nu\la}+ \del_\si^{\ \la}\ga^{\mu\nu}+ \del_\si^{\ \nu}\ga^{\la\mu} \right) ,
 \eea  
 the first term in~(\ref{OM}) can be expanded in terms of the K\"ahler form.  
 Pulling back this term to $Y$ results in 
 \be
 -\f{1}{(3!)^2 h}\ep^{ijk}\ep^{mnl} X^\mu_i X^\nu_j X^\la_k X^\rh_m 
 X^\si_n X^\del_l(\theta^\dagg \ga_{\mu\nu\la}\ga_{\rh\si\del}\theta)= 1  \, .
 \label{con}
 \ee

 Similarly, using 
 \[
 \ga_\rh\ga_{\mu\nu\la}= \f{i}{2}\ep_{\rh\mu\nu\la\et\del}\ga^{\et\del}\ga_7 
 +g_{\rh\mu}\ga_{\nu\la}+g_{\la\rh}\ga_{\mu\nu} + g_{\rh\nu}\ga_{\la\mu}\, ,
 \]
 and
 \bea
 \ep_{\rh\mu\nu\la\et\del}\, k^{\et\del}= 2(k_{\rh\mu}k_{\nu\la} + k_{\rh\nu}k_{\la\mu}
 + k_{\mu\nu}k_{\rh\la})\, , \nn 
 \eea
 the second term in~(\ref{OM})  can  be expanded in terms 
 of the K\"ahler form
 \bea 
 \f{1}{2}(\theta^\dagg\ga_{\mu\nu\la}
 \ga^\et\theta)(\theta^\dagg\ga_\et\ga_{\rh\si\del}\theta)&=& 
\nn\\&&\hspace{-2cm}
= k_{\la\nu}( g_{\mu\rh}k_{\si\del}
 + g_{\mu\del}k_{\rh\si} + g_{\mu\si}k_{\del\rh} 
 +i k_{\rh\mu}k_{\si\del} +i k_{\del\mu}k_{\rh\si}  
 + i k_{\mu\si}k_{\rh\del})+
 \nn \\
 &&\hspace{-2cm}\quad +k_{\nu\mu} ( g_{\rh\la}k_{\si\del}+ g_{\la\del}k_{\rh\si}
 + g_{\la\si}k_{\del\rh} 
 +i k_{\rh\la}k_{\si\del}+i k_{\del\la}k_{\rh\si}
 +i k_{\del\rh}k_{\si\la})+
 \nn \\
 &&\hspace{-2cm}\quad + k_{\mu\la} (g_{\rh\nu}k_{\si\del}
 + g_{\nu\del}k_{\rh\si}+ g_ {\si\nu}k_{\del\rh} 
 +i k_{\rh\nu}k_{\si\del}+ i k_{\del\nu}k_{\rh\si}
 +i k_{\si\nu}k_{\del\rh}) \nn \, .
 \eea
 It is now easy to see that
 \[
 \f{1}{2\cdot (3!)^2 h}\ep^{ijk}\ep^{mnl} X^\mu_i X^\nu_j X^\la_k X^\rh_m 
 X^\si_n X^\del_l (\theta^\dagg\ga_{\mu\nu\la}
 \ga^\et\theta)(\theta^\dagg\ga_\et\ga_{\rh\si\del}\theta)  
 = -\f{1}{2} {\sf k}_{ij}{\sf k}^{ij}\, .
 \]
 So finally we can write
 \bea
 \f{1}{2}s{\bar s} &=&\f{1}{2} \left(1 +\f{1}{3!\sqrt h}\ep^{ijk} X^\mu_i  X^\nu_j 
  X^\la_k \Om_{\mu\nu\la}\right) 
 \left(1 +\f{1}{3!\sqrt h}\ep^{mnl} X^\rh_m  X^\si_n  X^\del_l \Om^\dagg _{\rh\si\del}\right) \nn \\
 &=& 1 +\f{1}{2\cdot 3!\sqrt h}\ep^{ijk} X^\mu_i  X^\nu_j  X^\la_k (\Om_{\mu\nu\la} 
 + \Om^\dagg _{\mu\nu\la})- \nn \\ 
 &&- \f{1}{4}( X^\mu_i  X^\nu_j  k_{\mu\nu})(X^{i}_\rh X^j_\si k^{\rh\si}) 
 \nn \, ,
 \eea
 which is equation~(\ref{SAT}). Note that this equation puts a lower 
bound on the 
 induced volume of $Y$ 
 \[
 \int {\sqrt h}\, d^3\si  \geq -\f{1}{3!}\int \, X^*({\rm Re}\, \Om)\, ,
 \]  
 and indicates that this bound is saturated if and only 
 if $s={\sf k}_{ij}=0$.


\begin{thebibliography}{11}

 \bibitem{WSIG}
 E. Witten, \emph{Topological sigma models}, 
\cmp{118}{1988}{411}. 

\bibitem{BSV}
 M. Bershadsky, V. Sadov and C. Vafa, \emph{D-branes and topological field
   theory}, \npb{463}{1996}{420} [\hepth{9511222}].


\bibitem{WTFT}
 E. Witten, \emph{Topological quantum field theory},
\cmp{117}{1988}{353}.

\bibitem{WS}
E. Witten, \emph{Supersymmetric Yang-Mills theory on a four-manifold}, 
\jmp{35}{1994}{5101} [\hepth{9403195}].

\bibitem{WMO} 
E. Witten, \emph{Monopoles and four-manifolds}, \emph{Math.\ Research Letters} 
{1} (1994) 769 [\hepth{9411102}].

 \bibitem{MIR}
 E. Witten, \emph{Mirror manifolds and topological field theories}, 
 in \emph{Essays on mirror manifolds}, International press, 1992.

\bibitem{BBS}
 K. Becker, M. Becker and  A. Strominger, \emph{Fivebranes, membranes 
 and non-perturbative string theory}, \npb{456}{1995}{130} 
 [\hepth{9507158}].

\bibitem{HL}
 F.R. Harvey and H.B. Lawson, \emph{Calibrated geometries}, \emph{Acta Math.}\ 
{\bf 148} (1982) {47}.

\bibitem{DAVIS}
K. Davis, \emph{Generalized topological sigma model}, 
\hepth{9703113}.

\bibitem{MC}
 R. McLean, 
\emph{Deformations of calibrated submanifolds},  
\emph{Comm.\ Anal.\ Geom.}\  {\bf 6} (1998) 705
[\href{http://www.math.duke.edu/preprints/1996.html}
{Duke  preprint 96-01}].

 \bibitem{SYZ}
 A. Strominger, S.T. Yau and E. Zaslow, 
\emph{Mirror symmetry is T-duality}, 
\npb{479}{1996}{243} [\hepth{9606040}].

\bibitem{BAU}
 L. Baulieu and B. Grossman, \emph{Monopoles and topological field 
 theory}, \plb{214}{1988}{223}.

 \bibitem{JAJ}
 J.M. Figueroa-O'Farrill, A. Imaanpur and  J. McCarthy,  
 \emph{Supersymmetry and gauge theory on Calabi-Yau 3-folds},  
\plb{419}{1998}{167} [\hepth{9709178}].

\end{thebibliography}
\end{document}